\documentclass[10pt,onecolumn]{article}
\usepackage{times}
\usepackage[utf8x]{inputenc}
\usepackage[final]{pdfpages}
\usepackage{t1enc}
\usepackage{graphicx}
\usepackage{textcomp}
\usepackage[T1]{fontenc}
\usepackage{url}
\usepackage[ruled,vlined]{algorithm2e}
\usepackage{amssymb}
\usepackage{amsthm}

\newcommand{\Section}[1]{\vspace{-8pt}\section{\hskip -1em.~~#1}\vspace{-3pt}}

\newcommand{\edit}{}
\newcommand{\eedit}{}
\newcommand{\eeedit}{}
\newcommand\tstrut{\rule{0pt}{2.5ex}}

\begin{document}

\title{Analysis of close encounters with Ganymede and Callisto using a genetic n-body algorithm}
\author{Philip M. Winter \and Mattia A. Galiazzo \and Thomas I. Maindl \\ \small Department for Astrophysics, University of Vienna}
\onecolumn
\date{}
\maketitle

\begin{flushleft}
{\small Front. Astron. Space Sci., 22 May 2018, \url{https://doi.org/10.3389/fspas.2018.00016}}
\end{flushleft}

\section*{Abstract} 
In this work we describe a genetic algorithm which is used in order to study orbits of minor bodies in the frames of close encounters. We find that the algorithm in combination with standard orbital numerical integrators can be used as a good proxy for finding typical orbits of minor bodies in close encounters with planets and even their moons, saving a lot of computational time compared to \eedit{long-term} orbital numerical integrations. Here, we study close encounters of Centaurs with Callisto and Ganymede in particular. We also perform n-body numerical simulations for comparison. \edit{We find typical impact velocities to be between $v_{rel}=20[v_{esc}]$ and $v_{rel}=30[v_{esc}]$ for Ganymede and between $v_{rel}=25[v_{esc}]$ and $v_{rel}=35[v_{esc}]$ for Callisto.}\\
Keywords: Callisto, Ganymede, n-body, close encounters, genetic algorithm, celestial mechanics, numerical simulation, collisions

\Section{Introduction} 
Jupiter's large icy moons such as Ganymede and Callisto show countless impact craters across their surface. Studying these craters gives deep insights into the impactors as well as the moons themselves.
This is the first approach in the frame of future works of studying collisions with the outermost moon Callisto. We are especially interested in the Valhalla crater system, Callisto's biggest crater. This impact structure measures several hundred of kilometers in diameter and shows some extraordinary features such as an extensive ring system in the outskirts of the crater (Greeley et al., 2000; Stewart and Allen, 2002). It has been shown that studying the formation of the Valhalla crater reveals new insights regarding Callisto's subsurface composition (Winter et al., 2017).

In this first work we focus on the use of a genetic algorithm\footnote{herein `GA'} (which is described in Section 2 and in the Appendix for further details) as a proxy tool to find preliminary orbits of possible close encounters for bodies in the Solar System. In particular, we selected the Centaurs' population. Currently, 423 Centaurs are known\footnote{Data taken from JPL Small-Body Database Search Engine \url{https://ssd.jpl.nasa.gov/sbdb_query.cgi}}. It is estimated that the number of Centaurs with a diameter larger than 1\,km lies between $n\sim10^7$ (Volk and Malhotra, 2008) and about $n\sim8\cdot10^9$ (Di Sisto and Brunini, 2007; Fernández and Sosa, 2015; Napier et al., 2015). These bodies mainly origin from the Trans-Neptunian Objects and have very chaotic orbits. With semi-major axes between $a=5.5\/\mathrm{AU}$ and $a=30.1\/\mathrm{AU}$, they lie between the giant plantes, from which they are frequently ejected out of the Solar System via close encounters or even impact on one of the planets or their moons. Their lifetime is of the order of 10 Myr (see e.g., Horner et al., 2004a,b; Bailey and Malhotra, 2009; Galiazzo et al., 2016). A GA in combination with standard orbital numerical integrators can be used as a good proxy for typical orbits of minor bodies in close encounters or impacts with planets and even their moons, saving a lot of computational time compared to full orbital numerical integrations (Section 3.1). We also perform n-body numerical simulations (Section 3.2). We extrapolated the typical orbits in close encounters with their osculating elements and velocities (Section 3.1 and 3.2). This kind of orbital analysis can also be very useful for studying the origin of impactors in the last million years, i.e. the Bosumtwi impactor (Galiazzo et al., 2013). We particularly design the genetic algorithm to investigate close encounters with Callisto -- and for comparison -- Ganymede. Close encounters are the first natural approach to actual impact scenarios which we will study in following works.

\section{Methods}
Measuring close encounters or even collisions between minor and major bodies in the context of n-body simulations is computationally demanding. Typically one has to constrain the parameter space of the minor bodies to selected regions within the Solar System (e.g. Kuiper belt objects or specific families of objects). We use a genetic algorithm to find asteroids of the Centaur type family which are likely to have a close encounter with the Jovian moons Ganymede and Callisto. Centaurs mainly origin from TNOs and in particular from the Kuiper Belt (Galiazzo et al., 2016). The method is used to encounter the problem of a large parameter space of initial orbital elements (see Table 1). The genetic algorithm boosts the performance compared to classical searching grids by some orders of magnitudes within the given computation time. This allows for measuring a reasonable amount of datapoints to do a statistical analysis of typical close encounter families which marks the first step towards studying actual impact scenarios.

\subsection{Genetic Algorithms}
A genetic algorithm (Turing, 1950) is an iterative searching algorithm to find solutions for highly complex problems which can have large parameter spaces. GAs origin from the field of biology, therefore we may also use the corresponding terms. Further information is given in the Appendix.
GAs are efficient optimization methods for highly complicated functions. The population can overcome local minima quite easily either by the means of crossover or mutation. Moreover, GAs tend to find all possible families of solutions, even if they are unrealistic or uncommon.
However, GAs also have negative aspects. The implementation can be quite tricky and there is no general rule how to implement them efficiently because the functions (for fitness, crossover and mutation) and parameters (number of generations, number of individuals) have to be adapted to the problem as well as to the hardware. If the minimum fitness limit which is needed to solve the problem is not met (due to poor convergence or no learning process), the GA will not find any solutions.

\subsection{Genetic N-Body Algorithm}
\texttt{genbody} is a GA which is being developed to find \edit{close approach- and} collisional orbits of particles in the context of n-body simulations. Each generation of the GA corresponds to a distinct n-body simulation. During the simulation, the fitness of the population is measured. Afterwards, a new population is created by the means of crossover and mutation. \edit{A pseudo code of the genetic n-body algorithm is given in Algorithm 1.}

\begin{algorithm}[H]
\DontPrintSemicolon
initialize population\;
\For{g in generations}{
\While{$t < t_{end}$}{
take n-body step ($t \rightarrow t + h$)\;
measure fitness\;
}
crossover\;
mutation\;
}
\caption{genetic n-body algorithm}
\end{algorithm}

We use the code to find objects which are likely to collide with either Ganymede or Callisto within a certain time interval. The following list gives an overview how the GA and the n-body code are associated with each other. In analogy to section 2.1, we link the terms between the genetic algorithm and the actual problem:
\begin{itemize}
	\item population: Centaur type asteroids with random initial conditions
	\item DNA: initial Keplerian orbital elements ($a\/$, $e\/$, $i\/$, $\omega$, $\Omega$, $M\/$)
	\item generation: numerical simulation of the system via an n-body method
	\item evolution: iterative process of consecutive simulations
	\item fitness: score of each test particle at each simulation given by the fitness function
	\item crossover: combination of initial orbital elements of parent particles given by the crossover function
	\item child: a test particle with a new set of initial orbital elements
	\item mutation: small, random variation of initial orbital elements
\end{itemize}
The so-called fitness function is the function to be optimized. We use a fitness function of $f=1/{d_\mathrm{rel}}^2$ with $d_\mathrm{rel}$ being the minimum relative distance between a particle and the corresponding moon during each generation. \edit{The squared distance was found to be useful if the fitness is used as a probability distribution to sample quite fit parents for crossover, ensuring a high fraction of fit parents and thus increasing the overall performance of the GA.} Note that one can use other fitness functions to study completely different problems. One only has to find a quantity to score objects which show a specific behaviour to obtain a population which develops that behaviour. The fitness of each body is measured throughout the simulation. If a close approach with the moon occurs, a measurement is taken and the algorithm starts a completely new evolution process. If no close approach happens during the simulation, unfit objects are replaced by new children which are created by objects with a high fitness score. The initial orbital elements $y\in \{a,e,i,\omega,\Omega,M\}$ of the new children are exact 1:1 copies of their corresponding parent. Note that one could use other crossover functions which include two or more parents to create a new child. Next, a small mutation is applied to the whole population in order to avoid to get stuck in local maxima in the learning curve (generation vs. mean fitness). Since our n-body system turns out to be highly sensitive to the mutation rate $\chi$, it is crucial to find suitable values for each generation. A too small $\chi$ leads to a solution which gets stuck in a local maximum while a too large $\chi$ on the other hand leads to destruction of genetic information. This corresponds to random guessing without any learning process of subsequent generations of populations. Since the trajectories of small bodies in the Solar System are chaotic, the mutation has to be low enough to allow for similar trajectories of subsequent populations (except for the newborn child) within the simulation time. We therefore use an adaptive mutation rate $\chi$ which depends on the overall fitness evolution. \edit{A scaling factor $\zeta$ is introduced to control the learning curve by setting the amplitude of the mutation rate $\chi$: If the mean fitness between two successive generations varies too much, $\zeta$ is decreased to prevent a loss of DNA information. If the mean fitness between two successive generations varies too little (by less than 1\% in this work) or if the inidividuals of the population are getting too similar to each other, $\zeta$ is increased to ensure a healthy population. The mutation rate itself is given by the standard deviation $\sigma\/$ of each initial orbital element throughout the population (e.g. $\chi_e = \zeta \cdot \sigma_e$). The initial mutation scaling is set to $\zeta=0.1$ for all evolutions. At this point it should be noted that the functions for fitness, crossover and mutation are empirical functions found to yield good results (high performance due to learning curves with steep slopes) for this specific problem. Mathematical formulations of the functions we use for fitness, crossover and mutation are given in the Appendix.}

The size of the population is $n_\mathrm{pop} = 30$ and the total number of bodies in each simulation is $n_\mathrm{tot} = 38$, including the Sun, Venus, Earth, Mars, Jupiter, Saturn, Uranus and Neptune. The mass of Mercury was added to the Sun. The orbital elements for the massive bodies are obtained from the JPL HORIZONS system. \edit{We perform the simulations in the Jovian-centric system, as we expect only negligible changes of the results because the systematic error produced by the GA is significantly larger than the error made by not using a barycentric frame of reference. We use the Lie-Series n-body integrator as described in Hanslmeier
and Dvorak (1984) with a numerical accuracy of $\varepsilon=10^{-11}$. Note that we do not explicitly include Ganymede and Callisto in the simulations due to a performance gain. In contrast to symplectic n-body integrators, the stepsize of the Lie-integrator is limited by the minimum relative distance between any of the objects,
\begin{equation}
 h = \frac{1}{k_G} \left( \frac{\varepsilon \cdot \nu!}{\mathcal{D}} \right)^\frac{1}{\nu}
\end{equation}
where $h$ is the stepsize in days, $k_G$ is the Gaussian gravitational constant and $\varepsilon$ is the numerical accuracy. The quantities $\nu$ and $\mathcal{D}$ are obtained from the integrator itself during each step, where $\nu$ is associated with the number of used Lie-terms and $\mathcal{D}$ depends on the masses and relative distances between the objects (Eggl and Dvorak, 2010).} If we include the moons in the calculations, the stepsize would be limited by the distances within the Jupiter system \eedit{and much higher computational capacities would be needed in order to get the same amount of results}. Therefore, we use pseudo-moons which are characterized only by the distance to Jupiter without gravitation in order to be able to calculate distances. A measurement is taken if a particle has its closest approach within the Hill sphere of a moon. \eedit{The stepsize of the integrator during a Hill sphere crossing has to be low enough to ensure that the maximum spatial distance between two consecutive frames does not exceed the size of the Hill sphere. Otherwise, the particle would overleap the Hill sphere and no measurement would be detected. Therefore, an additional restriction for the upper bound of the stepsize is used. Moreover, if the moons are included in the simulations, their very short orbital periods would lead to significant stability issues with their orbits even at very low stepsizes. Note that since no actual positions of the moons are given during the simulations, the Hill spheres are represented by torus-shaped regions around Jupiter with their first radius being the relative distance to Jupiter and the second radius being the Hill radius of the respective moon. This approach can be used because of the statistical nature of the study, expecting a large number of measurements.} Since we are doing statistical studies of typical intersection velocities we found this simple approach to be highly effective.
Due to the nature of the GA, we are able to use a large parameter space for the random initial conditions of the test particles (see Table 1).
\begin{table}[htbp]
	\centering
	\begin{tabular}{lll}
		& min & max\\\hline
		$a\ [\mathrm{AU}]$\tstrut & 4.95 & 30.33\\
		$e\ [\mathrm{1.}]$ & 0 & 0.99\\
		$i\ [\mathrm{deg}]$ & 0 & 180\\
		$\omega\ [\mathrm{deg}]$ & 0 & 360\\
		$\Omega\ [\mathrm{deg}]$& 0 & 360\\
		$M\ [\mathrm{deg}]$& 0 & 360
	\end{tabular}
	\caption{Parameter space for random initial conditions of the test particles. The inclination is weighted by the density distribution of 420 known Centaurs \edit{to ensure a realistic distribution of objects.}}
\end{table}

We set the maximum simulation time $t_\mathrm{max}=165\,\mathrm{yr}$ in order to allow at least a full orbital period for each individual. The maximum number of generations $g\/$ is set to $g_\mathrm{max}=250$. \edit{A new iteration is started either if there is no measurement (i.e., close encounter with either Ganymede or Callisto) within these $g_\mathrm{max}$ iterations or as soon as such a close encounter is recorded.} Finally, for a quick check of the results with the GA method, we also check its output with the ones from full integrations of Centaurs (through all their lifetime) taking data kindly provided by Galiazzo et al. (2016), see section 3.2 for more descriptions. We compare these data with the orbits found for Ganymede, as far as close encounters with Jupiter were taken in account up to \edit{$d=0.01\,\mathrm{AU}$.}

\section{Results}
\subsection{Analysis of GA measurements}
After about 3300 CPU hours, total number of 531 and 625 measurements was obtained for Ganymede and Callisto, respectively. Each measurement corresponds to an individual evolution process of the GA and represents an individual close encounter scenario which contains all relevant information about positions, velocities, orbital elements, etc. about the bodies at both the time of closest approach and at the beginning of the respective simulation.

\begin{figure}[ht!]
	\centering
	\includegraphics[width=6cm]{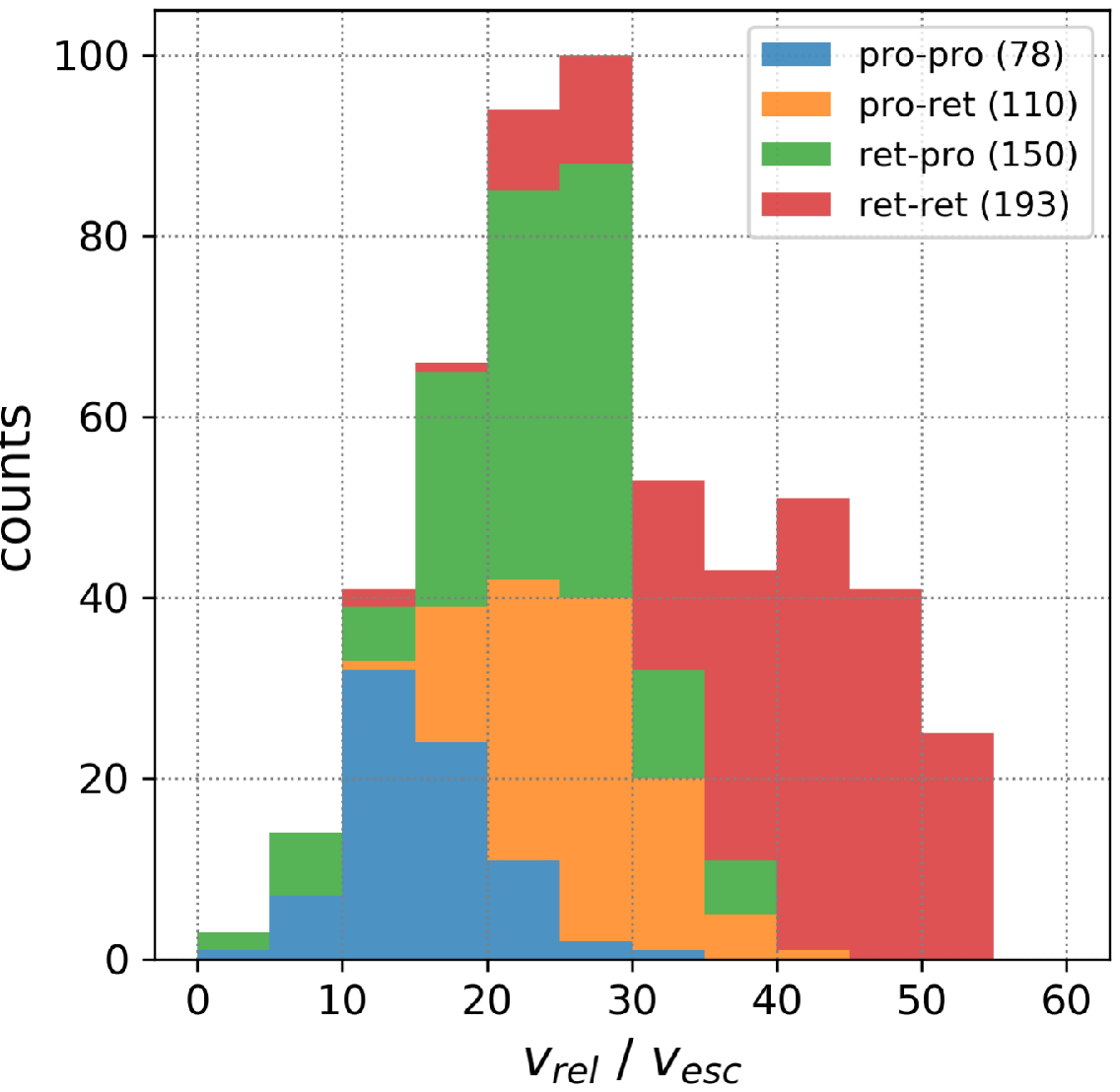}
	\includegraphics[width=6cm]{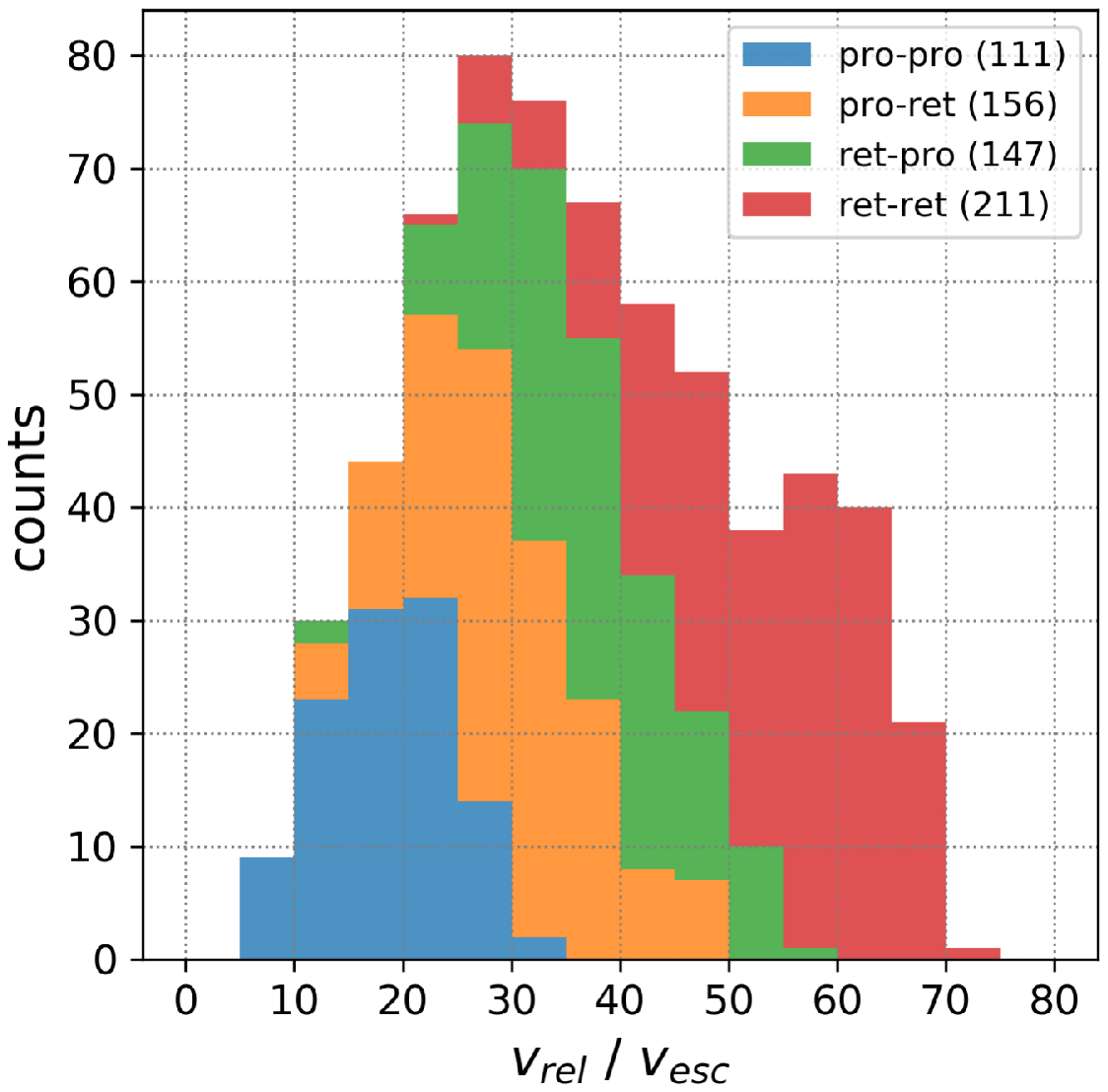}
	\caption{left: Ganymede, right: Callisto. The histograms show the relative velocities between the test particles and the corresponding moon at closest approach in units of the escape velocity. The four colourized components represent the possible geometrical encounter scenarios. For example, the orange component \textit{pro-ret} refers to a particle having an inclination smaller than $i=90^\circ$ (therefore prograde), but encountering the moon retrograde with respect to the Jovian System. The numbers in parentheses correspond to the respective number of measured datapoints. All bins are stacked.}
\end{figure}

Figure~\ref{vel} presents the main results of these measurements, showing the relative velocities at closest approach with respect to the moons. We can split the data into 4 classes, depending on the geometrical properties of the particle trajectories:

\begin{enumerate}
 \item \edit{The low-velocity class (\textit{pro-pro} encounters)}
 \item \edit{The intermediate-velocity class A (\textit{pro-ret} encounters)}
 \item \edit{The intermediate-velocity class B (\textit{ret-pro} encounters)}
 \item \edit{The high-velocity class (\textit{ret-ret} encounters)}
\end{enumerate}

The terms \textit{pro-pro}, \textit{pro-ret}, \textit{ret-pro} and \textit{ret-ret} refer to the corresponding geometries, where the first part describes the prograde or retrograde movement with respect to the Jovian System (with other words: $i<90^\circ$ is \textit{pro} and $i>90^\circ$ is \textit{ret}) and the second part describes the direction of flight with respect to the corresponding moon within the Jovian System, moving parallel (\textit{pro}) or antiparallel (\textit{ret}) to the respective moon. More detailed statistics of the four classes are given in Table 2.

\begin{table}[ht!]
\centering
\begin{tabular}{llllll}
Ganymede & $\mu\/[\mathrm{km / s}]$ & $\sigma\/[\mathrm{km / s}]$ & $\mu\/[\mathrm{v_{esc}}]$ & $\sigma\/[\mathrm{v_{esc}}]$ & $n$ \\ \hline
class 1, \textit{pro-pro} \tstrut& 15.03 & 4.55 & 15.36 & 5.17 & 78 \\
class 2, \textit{pro-ret} & 24.38 & 3.09 & 26.00 & 5.33 & 110 \\
class 3, \textit{ret-pro} & 22.39 & 5.59 & 23.45 & 6.68 & 150 \\
class 4, \textit{ret-ret} & 39.29 & 3.84 & 40.65 & 8.48 & 193 \\ \\
Callisto & $\mu\/[\mathrm{km / s}]$ & $\sigma\/[\mathrm{km / s}]$ & $\mu\/[\mathrm{v_{esc}}]$ & $\sigma\/[\mathrm{v_{esc}}]$ & $n$ \\ \hline
class 1, \textit{pro-pro} \tstrut& 11.63 & 3.40 & 18.44 & 5.96 & 111 \\
class 2, \textit{pro-ret} & 19.41 & 3.57 & 29.53 & 7.91 & 156 \\
class 3, \textit{ret-pro} & 23.85 & 3.59 & 36.84 & 8.40 & 147 \\
class 4, \textit{ret-ret} & 34.96 & 3.12 & 52.86 & 10.45 & 211 \\ \\
\end{tabular}
\caption{\edit{Mean and standard deviations of close encounter velocities for the four classes, as well as their numbercounts.}}
\end{table}

\edit{The most probable close encounter velocities can be seen between $v_{rel}=20[v_{esc}]$ and $v_{rel}=30[v_{esc}]$ for Ganymede and between $v_{rel}=25[v_{esc}]$ and $v_{rel}=35[v_{esc}]$ for Callisto.}

\edit{Further observations can be obtained from the results:}
\begin{itemize}
 \item \edit{The overall form of the relative velocity histograms can be reproduced by overlapping Gaussian distributions which are represented by the four classes.}
 \item \edit{The classes are overlapping stronger for Ganymede than for Callisto, even swapping places (in velocity) when comparing class 2 and class 3 for Ganymede.}
 \item \edit{There is a clear trend favouring retrograde encounters (for both Jupiter and the respective moon), with most close encounters being \textit{ret-ret} (36.3\% for Ganymede, 33.8\% for Callisto).}
\end{itemize}

\begin{figure}[ht]
	\centering
	\includegraphics[width=6cm, height=6cm]{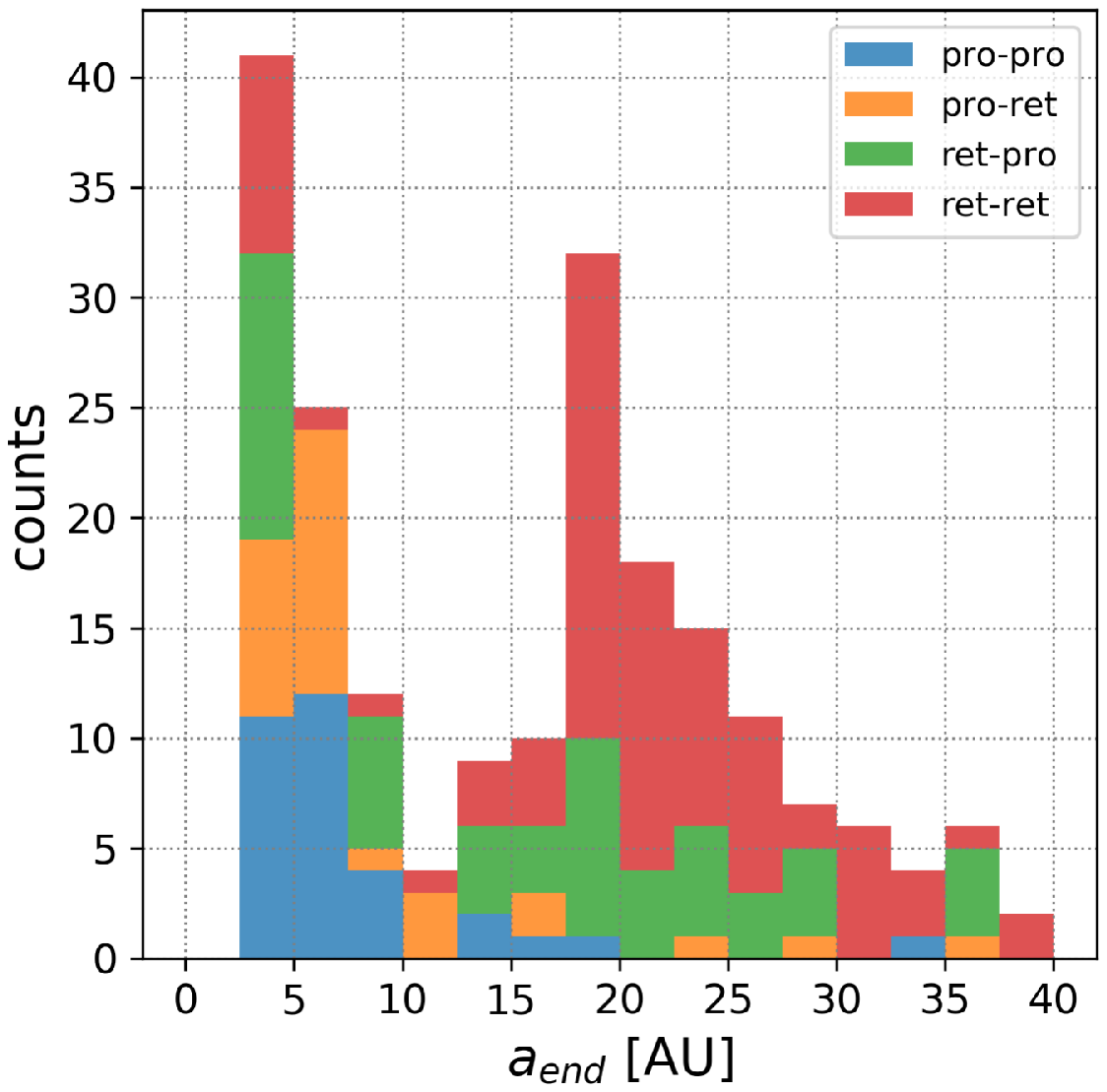}
	\includegraphics[width=6cm, height=6cm]{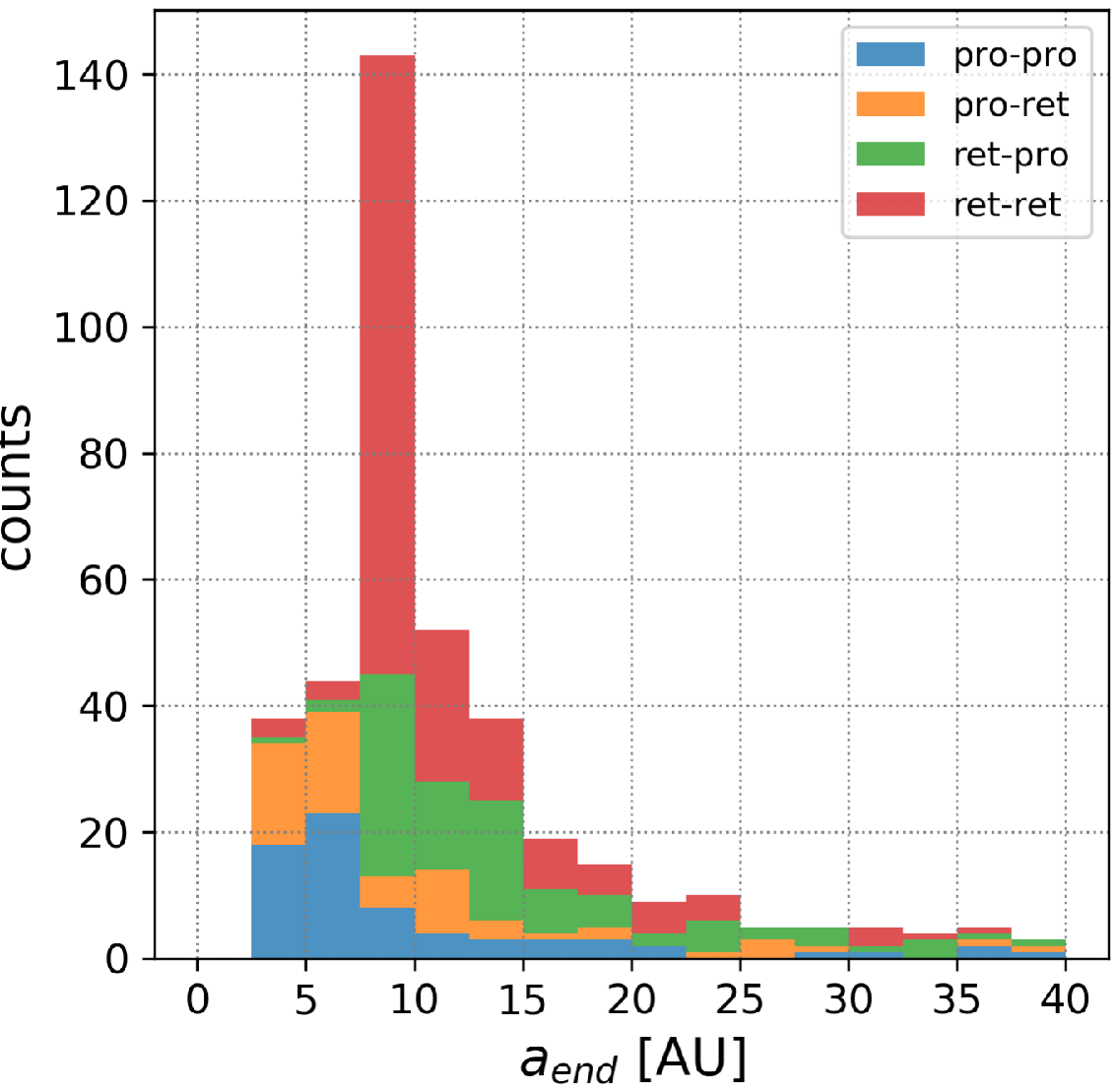}
	\includegraphics[width=6cm, height=6cm]{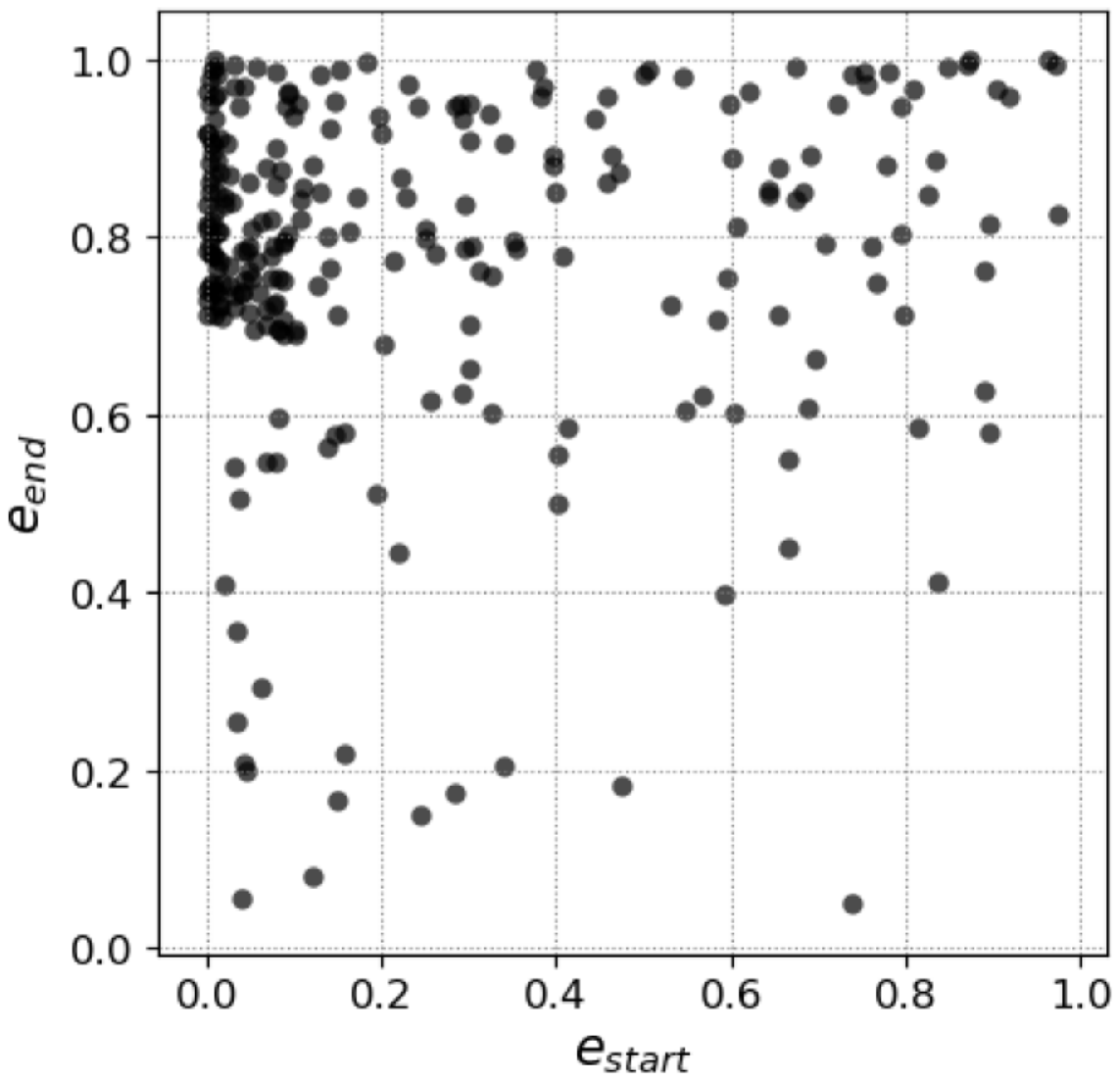}
	\includegraphics[width=6cm, height=6cm]{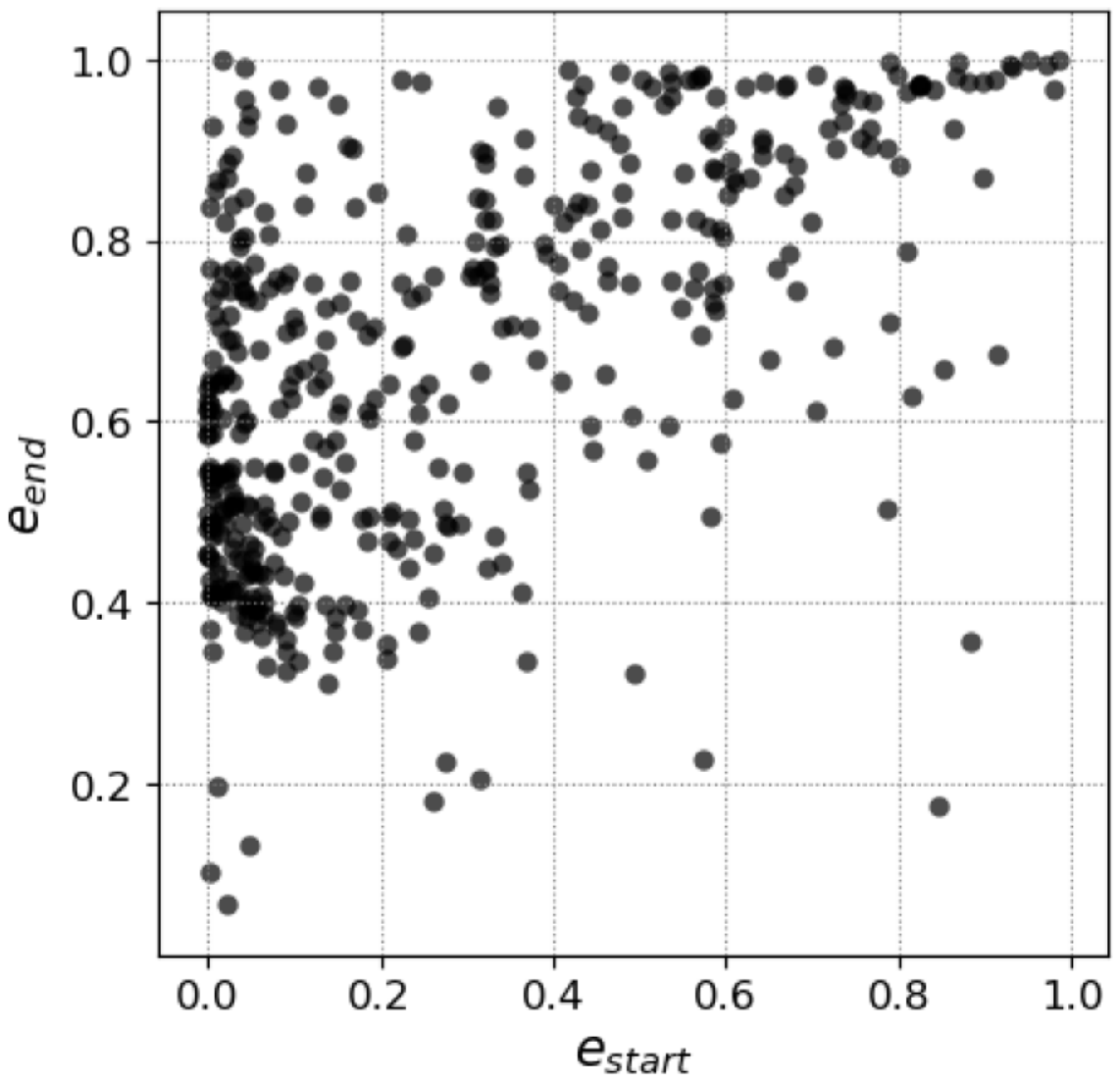}
	\includegraphics[width=6cm, height=6cm]{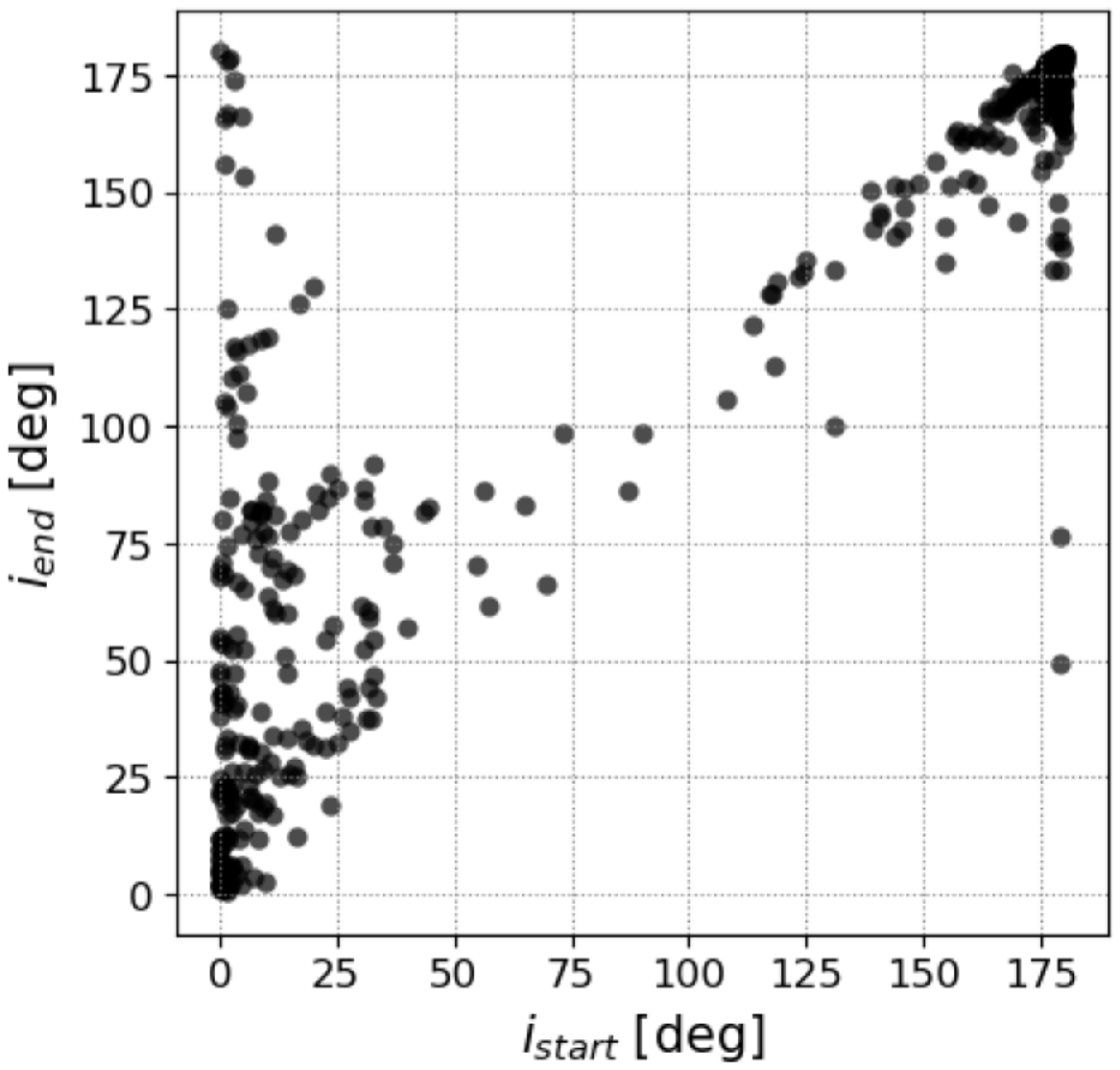}
	\includegraphics[width=6cm, height=6cm]{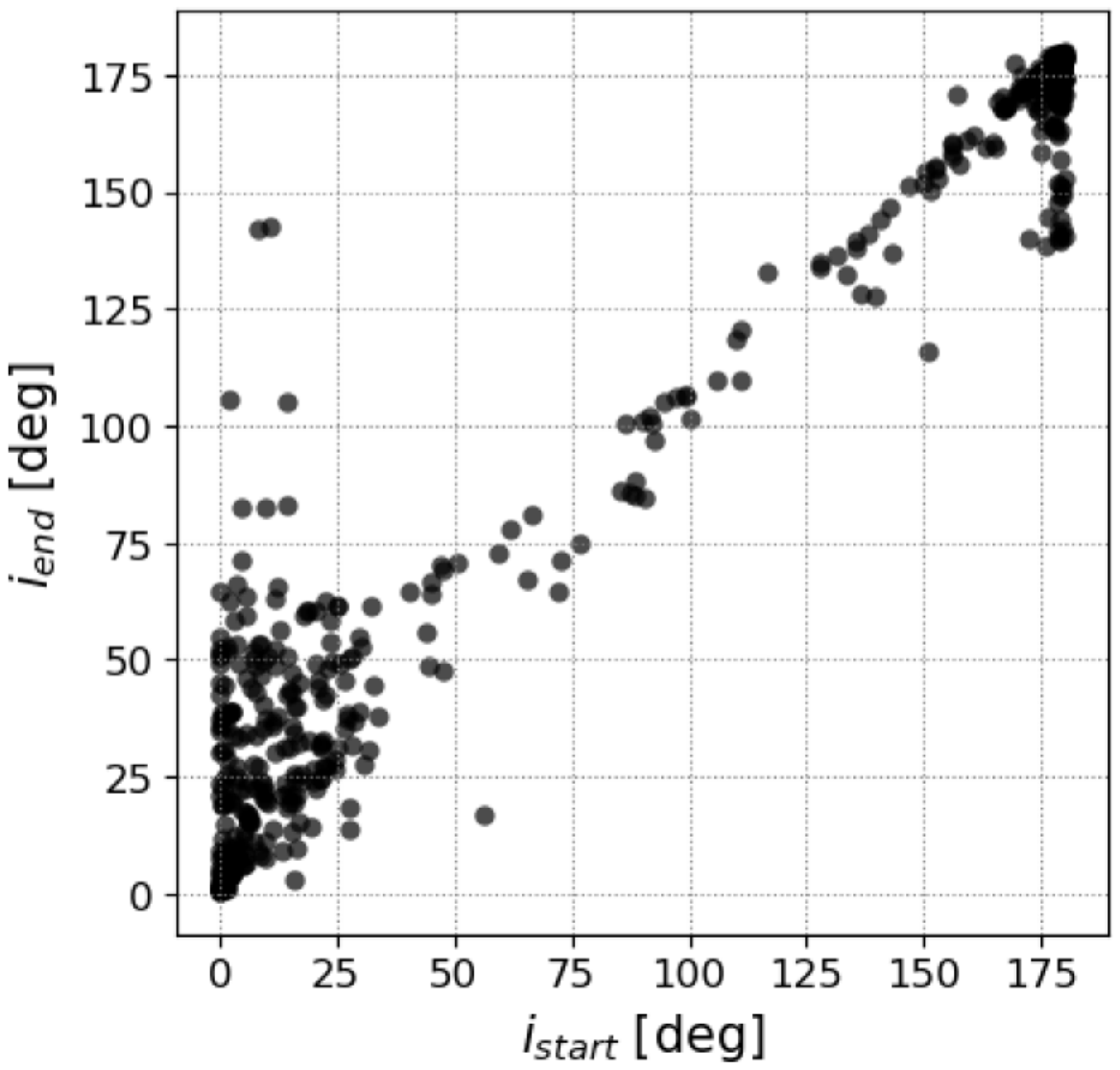}
	\caption{Left column: Ganymede, right column: Callisto. The top row shows the semi-major axes at closest approach. The middle row shows the initial eccentricity of the last generation versus the eccentricity at the measurement. The bottom row shows the respective inclinations. Each datapoint corresponds to an independent evolution.}
\end{figure}

The first row in Figure 2 shows the final semi-major axes at the time of the closest approach. While Ganymede shows two distinct peaks at $a\simeq 5\ \mathrm{AU}$ and $a\simeq 20\ \mathrm{AU}$, Callisto only has one single, large peak at $a\simeq 9\ \mathrm{AU}$.
The other panels shows the respective quantity at starting time of the last generation versus the same quantity at the time of the respective measurement. A 1:1 relation corresponds to no change of the orbit during the simulation, while a large scattering means significant changes.
As we can see for the eccentricities, there is a significant scattering towards larger values. The plots imply that the scattering for Ganymede close encounters is higher than for Callisto. More than half of the particles even undergo the transition from elliptic \edit{($e<1$)} to hyperbolic \edit{($e>1$)} trajectories. The strong scattering is also apparent in the semi-major axes.
The inclination data shows only minor scattering between $i=50^\circ$ and $i=150^\circ$, which implies quite stable configurations within the simulation time. However, at low inclinations particles are scattered over the full parameter space and prograde ones tend to get scattered stronger than their retrograde counterparts.

Several other obervations can be obtained from the datasets:
\begin{enumerate}
	\item The needed number of generations for taking the measurement \edit{grows steadily until about $g=100$} before falling again rapidly.
	\item The most dramatic changes of initial orbits happen during the first few tens of generations. 
	\item It is easier for the GA to find intersection orbits after a short simulation time. Therefore, the time of measurement peaks towards low values \edit{with a mean simulation time of $t_{mean}=45.5\,\mathrm{yr}$.}
	\item The intersection probability is approximately twice as high for Callisto because her Hill radius is larger than Ganymede's. For statistical reasons, we therefore assigned more computation time to Ganymede.
	\item Two actual collisions are measured for Ganymede (with impact velocities of $v_\mathrm{rel}=34.6\ \mathrm{km/s}$ and $v_\mathrm{rel}=42.2\ \mathrm{km/s}$, respectively), none for Callisto.
\end{enumerate}

\subsection{Preliminary Parent Bodies' Orbits}
We take the orbital evolution of Centaurs to do a comparison between the predicted Centaurs' orbits at close encounters and the Jovian moons. We integrate forward a sub-sample of the Centaurs for 30 Myrs and check all the close encounters with Jupiter, using the Lie-integrator. \edit{This study considers only close encounters up to a distance of $d=0.01\,\mathrm{AU}$ from Jupiter similarly to Galiazzo et al. (2016), but with an encounter radius 4 times smaller at the cost of computational time. Thus, the comparison is limited to Ganymede only (since he has a distance of $d\simeq0.0072\,\mathrm{AU}$ to Jupiter. Callisto has a distance of about $d\simeq0.0126\,\mathrm{AU}$ to Jupiter.)\footnote{We assume the semi-major axes of the moons as a proxy for the distance, neglecting their small eccetricities.}. 

\eedit{We take 319 Centaurs with 15 clones in each interval ranging over $5\,\mathrm{AU}$ in semi-major axis (for a total of 5104 bodies)}\footnote{The first region is the one with a semi-major axis between 5\,AU and 10\,AU. The second region is between 10\,AU and 15\,AU and congruently for the other regions up to 30\,AU.}} in order to quickly get a statistical sample which covers the entire Centaur region from 5\,AU to 30\,AU in semi-major axis. This approach is sufficient to give an idea of these kinds of bodies approaching Jupiter and its moons and to have a compareable sample to the orbits produced by the GA.

\edit{From the evolution of \eedit{5104} objects, a total number of \eedit{292} measurements was obtained for Ganymede close encounters.} From our sample of Centaurs we find that $\sim 22.6$\% can have close encounters with Jupiter. As the percentage of Centaurs which can cross Ganymede orbits is about 20.1\% (8.7\% with $e<1$), almost all the Centaurs' close encounters with Jupiter ($\sim 89$\%) can reach Ganymede's orbit in the range of its Hill sphere. Figure 3 shows the comparison between GA and Lie integrations for the respective orbital sample output in the Hill sphere of Ganymede.
\begin{figure}[ht!]
	\centering
	\includegraphics[width=13cm]{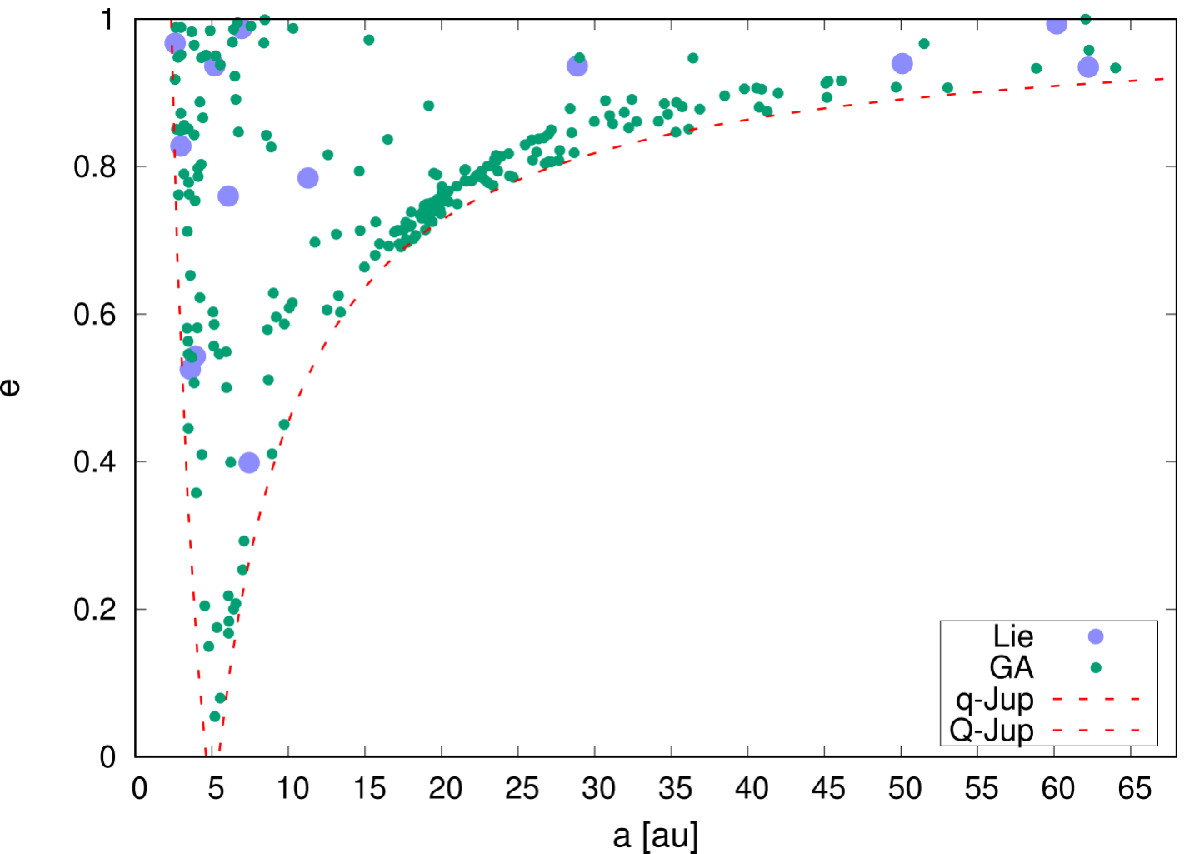}
	\includegraphics[width=13cm]{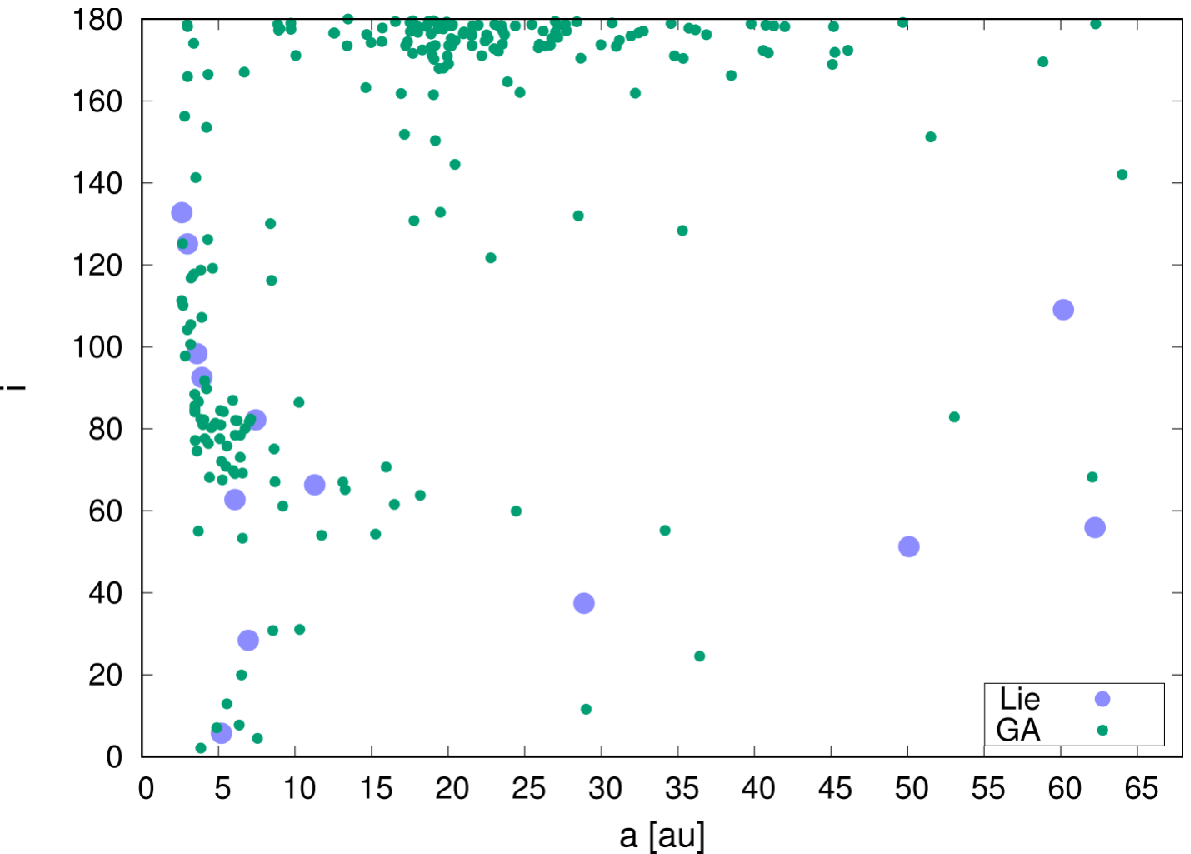}
	\caption{\edit{Comparison between measurements of GA and Lie integrations of evoluted orbits for Ganymede close encounters. For the GA, all measurements are taken within Ganymede's Hill sphere. For the Lie integrations, the measurements are taken as described in the text.}}
\end{figure}

\eeedit{Figure 3 suggests that the datasets are in quite good agreement, with the GA and the Lie datapoints covering a similar area in the $a-e$ phase space. However, a significant sample at high inclinations is apparent in the GA case, which makes the GA measurements not fully consistent with the measurements obtained by the Lie-integrations. An explanation for this effect is given in section 4. It turns out that these high-inclination datapoints are also responsible for the dense cloud in the $a-e$ phase space between $a=15-30\,\mathrm{AU}$ and $e=0.7-0.8$. Otherwise, the distributions in the $a-i$ space are reasonably in agreement, especially at semi-major axes below $a=15\,\mathrm{AU}$. Even the major cloud between $a=2.5-15\,\mathrm{AU}$ and $i=50-140^\circ$ is reflected well. In order to quantify the resemblence of the two distributions obtained via the GA and Lie-integrations, we deploy a two-dimensional Kolmogorov-Smirnov test (Press and Teukolsky, 1988). Due to the reasons mentioned in section 4 we exclude the non-physical high-inclination datapoints with $i>160^\circ$. For the $a-e$ distributions we get a p-value of $p=0.27$, for the more diverse $a-i$ distributions $p=0.058$. While the $a-e$ distributions do not differ significantly, the $a-i$ distributions show significantly less correlation, which indicates the need for a refined GA method as laid out in section 5.}

\section{Conclusions}
\edit{The velocities of the four classes as described in section 3.1 can be explained by geometrical considerations: Class 1 encounters happen if the particles experience deceleration from the Jupiter flyby and additionally 'lose' relative velocity due to the parallel direction of flight compared to the respective moon. The intermediate classes 2 and 3 either experience deceleration or acceleration from the Jupiter flyby and additionally 'gain' or 'lose' relative velocity due to the antiparallel or parallel direction of flight compared to the respective moon. Both acceleration by Jupiter and a 'gain' of relative velocity is true for class 4.}

\edit{Comparing the numbercounts of the classes, a correlation exists with higher fractions of both types of retrograde encounters being more probable. The higher fractions of retrograde encounters can be described by a simple geometrical effect: While a particle moves within the torus-shaped Hill region (the complete area which is accessed by the moon's Hill sphere during a full orbital period), the probability for a retrograde measurement is much higher than a prograde one since the typical relative velocity between the particle and the moon is high. This finding supports the observation of the heavily cratered front-side of the rotationally bound moons. For example, many large crater systems on Callisto (namely Valhalla, Asgard, Adlinda, Utgard, etc.) are located at the front-side.}

\edit{Comparing class 2 and class 3, it seems that the influence caused by the flyby at Jupiter and the parallel or antiparallel direction of flight are similar in strength, producing similar numbercounts for these classes. More discreet or weaker effects like the selection of the parameter space and the behaviour of the GA may influence the shape of the curves as well as the numbercounts for the classes. However, the overall shape of the individual components can be interpreted as Gaussian distributions, covering a large interval for possible close encounter velocities.}

\edit{Recalling the distributions of semi-major axes in Figure 2, the second peak for Ganymede is caused by the stronger scattering of the classes \textit{ret-pro} and \textit{ret-ret}. This means that Jupiter scatters retrograde orbits stronger than prograde orbits at smaller distances to Jupiter.} Many particles end up with high eccentricities, regardless of their initial values. Note that the simulation time of $t_\mathrm{max}=165\,\mathrm{yr}$ indicates that these drastic changes are caused by a single or only a handful of close encounters with Jupiter. The stronger scattering towards higher values of the semi-major axes and eccentricities for Ganymede is intuitive, because particles experience a stronger acceleration by Jupiter during the flyby. The data implies that more than half of all particles even undergo the transition from eccentric to hyperbolic trajectories during the simulations (taking into account we only considered elliptic orbits initially). The comparison between the final semi-major axes indicates that the large peak (at Callisto) vanishes at closer distances to Jupiter (towards Ganymede) and gets scattered over a wider range at typically higher values.

\eedit{For the inclination, the GA has a selection effect for the sector $a>15\,\mathrm{AU}$ and $i>160^\circ$, although the initial setup for the first generation resembles a realistic inclination distribution. This is clearly a selection bias of the GA, as it forces the population to settle either in very low or very high inclination orbits. This settling is due to the fact that statistically the relative distances between the test particles and the Jupiter system are smaller if their orbital planes are aligned. This leads to higher fitness values at very low/high inclinations. However, as a first approximation the GA works well as a predictor for close encounters for prograde orbits. Therefore, we recommend using the GA only for prograde orbits at current state.}

The comparison with the numerical integations (see Figure 3) reveals that the GA represents the overall close encounter situation quite well in the $a-e$ space and pretty well in the $a-i$ space when excluding retrograde orbits with very high inclinations. Both density distributions cover similar areas in the parameter spaces with almost all datapoints from the Lie measurements having their counterparts in the GA measurements. \eeedit{The statistical analysis via a Kolmogorov-Smirnov test even suggests a correlation between the density distributions for the $a-e$ phase space.} Note that the simulations in the GA have a very short integration time of $t_\mathrm{max}=165\,\mathrm{yr}$, which do not allow for modelling the long-term behaviour of chaotic orbits. \eedit{However, the GA still yields a good coverage of parameter space, even with this short simulation time}. Interestingly, the GA also finds orbits which lie far beyond the initial parameter space of semi-major axes given in Table 1. This indicates, that the implementation of the mutation function of the GA is flexible enough to explore the parameter space beyond its limits. The $a-i$ parameter space is reproduced for inclinations up to $i=160^\circ$, \eeedit{although the statistical analysis suggests no significant correlation in density}. As already stated, the high number of retrograde measurements is a selection bias of the GA.

For future work with the GA we may include the simulation time and further properties of the orbits in the fitness function in order to avoid too short simulations and make high inclinations less likely, for example. We may also include the TNO region as an important extension to the parameter space.

The GA can easily be adapted in order to efficiently measure actual collisions with any given object. In this work we measure only two collisions with Ganymede because as soon as the first particle has its closest approach within the Hill sphere, a completely new evolution process is started.

\edit{In summary, the results from the GA are not fully consistent compared to the classical approach because the underlying principles are different. GAs in general tend to find all possible solutions to a given problem rather than only the realistic, physical ones. However, in our implementation this effect can be overcome by optimizing the GA either to avoid clearly non-physical solutions or to enhance realistic solutions by increasing the simulation time, enlarging the parameter space and refining the functions for fitness, crossover and mutation.}

\section{Discussion}
\edit{The comparison with the Lie integrations reveal that the genetic n-body algorithm yield both a high number of physical as well as non-physical results. For example, an unrealistic high number of retrograde orbits are found while using a realistic probability distribution for the random initial inclinations. This is clearly a selection effect of the GA, as it finds that the probability for measuring retrograde encounters is significantly higher than for prograde ones. Several reasons such as a too short simulation time, too powerful crossover- and mutation functions or the choice of hyperparameters (such as the population size, number of generations, etc.) can be responsible for causing this high fraction non-physical results. However, it is expected that this high fraction can be significantly reduced by applying one or more of the following improvements:}

\begin{itemize}
 \item \edit{A higher simulation time for each generation enables more dynamical effects in general. Therefore, the GA will also tend to produce more physically motivated results.}
 \item \edit{The fitness function can be refined to avoid solutions which are clearly non-physical. For example, one may introduce additional terms which depend on the inclination, the time of measurement, etc.}
 \item \edit{Since crossover tends to find orbits within the initial random parameter space, this parameter space should be large enough, e.g. 50\% larger compared to the parameter space of interest. Regions outside the inital random parameter space are only accessible via mutation.}
 \item \edit{We found the mutation function to have a significant effect on the behaviour of the GA's learning curve (generation vs. mean fitness). An optimized mutation function can boost the overall performance of the GA drastically, enabling the use of a higher simulation time, a larger population, obtaining more measurements, etc. with the same computation resources.}
 \item \edit{A larger population is able to cover the parameter space more homogeneously and may reveal further close encounter families with low probabilities.}
 \item \edit{Like in this work, the non-physical results can be efficiently filtered by comparing the results with classical approaches.}
\end{itemize}
\edit{However, the GA also yields useable results, especially on the small-scale close encounter dynamics. We find it to be an efficient tool to get a rough idea of the underlying dynamics of the problem and the expected families of solutions before investigating into more detailed analysis with classical approaches. The GA supports the use of existing approaches rather than replacing them. In this work, the GA efficiently finds all possible close encounter geometries even beyond the initial parameter space with low computational effort. \eedit{The measurements cover all major areas of the parameter spaces in semi-major axis, eccentricity and inclination. Even a weak correlation in the density distributions \eeedit{for the $a-e$ space} is apparent when comparing the results of the GA with the long-term Lie integrations. This is quite interesting, as the used methods represent completely different approaches to the underlying problem of close encounters.} This encourages more detailed studies with optimized algorithms. In theory, the GA can be used to study a variety of different problems in celestial mechanics, given the appropriate fitness function and adapted functions for crossover and mutation.}

\edit{Apart from the behaviour of the GA itself, relevant information is obtained from the measurements:}
\begin{itemize}
 \item \edit{The four classes, which are motivated by geometrical considerations, can be distinguished well in the datasets. The classes allow for a more detailed and structured way for analysing close encounter events in the Jovian System.}
 \item \edit{One may distinguish between different impact scenarios depending on the impact velocity. For example, there is no need for analysing retrograde collisions if the particle is classified as class 1 and vice-versa for prograde collisions and class 4.}
 \item \edit{There are significantly more retrograde than prograde encounters for both moons. This fact is supported by the heavily cratered front-side of the rotationally bound moons.}
 \item \edit{As shown in Figure 2, most of the particles get scattered quite drastically in eccentricity during the close encounter. A high fraction may even end up in hyperbolic trajectories.}
 \item \edit{Moreover, the distributions of semi-major axes reveal a double-peak structure for Ganymede in contrast to a single-peak structure for Callisto. This can be explained by a stronger scattering of the classes \textit{ret-pro} and \textit{ret-ret} at smaller distances to Jupiter.}
\end{itemize}

\section*{Author Contributions}
PW wrote the code for the genetic algorithm and the transformations for the osculating elements. He wrote about 70\% of the paper, especially the part inherent to the genetic algorithm, the core of this paper. MG has contributed performing the full orbital numerical part section, selecting the initial populations for this study, he suggested how to consider the close encounters with the Jupiter moons and he edited several parts of the paper for quality improvements in any section of the paper, apart the description of the genetic algorithm. He wrote about 15\% of the paper. TM helped in editing the quality of the paper and wrote about 15\% of the paper. He also helped in the description of the genetic algorithm and performing the Kolmogorov-Smirnov test.

\section*{Acknowledgements}
The authors wish to thank the three anonymous referees for their comments which helped to significantly improve the manuscript. The authors acknowledge support by the FWF Austrian Science Fund projects S11603-N16 (PMW and TIM) and P23810-N16 (MAG), respectively.

\section*{Appendix}
The principle of a genetic algorithm can be described in a three-step process:
\begin{enumerate}
	\item The first generation is initialized in the system. The initial DNA of the individuals should cover the full parameter space in which at least one solution may exist. If there is no solution within the parameter space, no solution may be found by the GA. Note that the final solution of the problem is not only affected by the DNA, but also by other factors like the dynamics of the system itself, the fitness function, the crossover function, the mutation function, and the number of generations.
	\item The population evolves within the system and the fitness of each individual is measured. Usually this is the part which is computationally most demanding. It is crucial to score each individual correctly with the fitness function to enhance finding the final solution to the given problem.
	\item At the end of each generation the individuals with the least fitness die out and are replaced by new individuals (children) via crossover. Moreover, a small mutation is applied to the whole population. We then have a new population using the DNA of the previous one. Steps 2 and 3 are repeated until the fitness of the population reaches a certain level, which corresponds to the solution of the problem. At least one individual has to die out and there must be at least one new child at each generation. The number of children has to be equal to the number of dying individuals in order to keep the total number of individuals in the population equal for all generations. Note that each new generation is likely to have a higher mean fitness than the previous one which can be interpreted as a learning process.
\end{enumerate}
\subsection*{Population}
The number of individuals in the population is crucial for a GA to work. Choosing a too low number, the individuals tend to become all equal after a few generations. Since they are all equal, the learning process can only take place via mutation, which is often insufficient to solve the problem properly. This situation significantly hinders finding solutions within limited computation time. The population may also get stuck in local maxima (of the fitness curve) easily and no good solution will be found.

If the number of individuals is too high on the other hand, many generations are needed to observe a learning effect. One may also encounter performance problems because each individual has to evolve in the system at each generation. However, in principle it is good to prefer a rather high number of individuals for several reasons such as coverage of the parameter space, minimizing the risk of getting stuck in local maxima, and an overall healthier population of competing families using different approaches to solve the given problem. Considering these facts, one may choose an individual number as high as possible for given computation power and memory space. Note that for GAs it is often relatively simple to implement parallel computing with either multiple CPUs or with GPUs. A higher number of individuals may then be used.
\subsection*{DNA}
The properties of each individual are summarized in their DNA. For example, the DNA of an asteroid may be its orbital parameters as well as its mass, shape, albedo, composition, etc. The DNA depends on the nature of the problem itself. In general it is good to have as much relevant information in the DNA as possible to include as many possible aspects of the problem.
\subsection*{Fitness Function}
The fitness function scores each individual by its behaviour (which itself is based on its DNA) within the system. The better an individual can solve the given problem, the higher the fitness should get. Keep in mind that the problem to be solved has to be imprinted in the fitness function. Since we want to analyse close encounters, our fitness function is given by the inverse of the squared relative distance between a particle and the corresponding Jovian moon. \edit{Algorithm 2 shows the implementation used for the genetic n-body algorithm. The fitness for each individual in one generation is given by the maximum fitness at any timestep of the n-body integration.}\\
\begin{algorithm}[H]
\DontPrintSemicolon
\For{i in individuals}{
$d_{rel, i}= || \vec{x_m} - \vec{x_i} ||$\;
$f = d_{rel, i}^{-2}$\;
$f_i = max(f_i, f)$\;
}
\caption{measure fitness}
\end{algorithm}
\subsection*{Generations and Evolution}
During the iterative searching process (evolution), the population ideally gets better and better with each generation to find good solutions to the problem. Nevertheless, there may be generations with lower fitness than the previous ones. If the population settles in a local maximum, it has to overcome it again. This means, that at least one individual has to find a higher maximum to guide other individuals towards the respective direction in the parameter space. Then, during the process of overcoming, the mean fitness of the population may even decrease for a short period before rising again to new maximum values.
\subsection*{Natural Selection and Crossover}
Natural selection is another key component of a GA. Relatively fit `parents' are allowed to pass their DNA to the following generation by creating a new `child'. The crossover function can take many different forms, as long as it conserves and incorporates the parents' DNA. In many cases, a simple crossover function (mean values between parent A and parent B) is sufficient to produce children which are likely to have a higher fitness than the mean fitness of the population. Typically, the least fit individuals are replaced by new children to keep the population size constant. The crossover and mutation functions have to be adjusted to each other to ensure a steep learning curve. If the crossover is too strong, the population gets stuck in local maxima and mutation is not sufficient to overcome them again. If the mutation is too strong, the information of the DNA gets destroyed too fast and no learning process can happen. The fitness of the population may even decrease and settle at a very low value. \edit{Algorithm 3 shows the implementation used for the genetic n-body algorithm. The vector y contains the 6 initial orbital elements for the respective object. A 1:1 copy is being made using a parent with high fitness.}\\
\begin{algorithm}[H]
\DontPrintSemicolon
\For{ch in children}{
choose parent pa: $f_{pa} > f_{ch}$\;
$\vec{y_{ch}} = \vec{y_{pa}}$\;
}
\caption{crossover}
\end{algorithm}
\subsection*{Mutation Function}
The mutation function is used to apply small, random variations to the DNA of every individual after every generation. These variations are needed in order to overcome local maxima and to keep the population `healthy'. This means, that the DNA of the individuals can not degenerate to the same values for the whole population where no learning process can happen any more. \edit{Algorithm 4 shows the implementation used for the genetic n-body algorithm. Each orbital element of each individual is mutated separately using random numbers between 0 and 1. The scaling factor $\zeta$ controls the amplitude of the mutation rate $\chi$.}\\
\begin{algorithm}[H]
\DontPrintSemicolon
\If{$\overline{f_t} > 20 \cdot \overline{f_{t-h}}$ or $\overline{f_t} < 0.2 \cdot \overline{f_{max}}$}{
$\zeta = \zeta \cdot 0.8$\;
}
\If{$(\overline{f_t} > 0.99 \cdot \overline{f_{t-h}}$ and $\overline{f_t} < 1.01 \cdot \overline{f_{t-h}})$ or $max(\vec{f_t}) / \overline{f_t} < 2$}{
$\zeta = \zeta \cdot 2$\;
}
calculate $\sigma_a$, $\sigma_e$, $\sigma_i$, $\sigma_\omega$, $\sigma_\Omega$, $\sigma_M$\;
$\vec{\chi}=\zeta \cdot \vec{\sigma}$\;
\For{i in individuals}{
$\vec{y_i} = \vec{y_i} + random \cdot \vec{\chi}$\;
}
\caption{mutation}
\end{algorithm}

\end{document}